\documentclass[showpacs,prl]{revtex4}
\usepackage{graphicx}% Include figure files
\usepackage{dcolumn}% Align table columns on decimal point
\usepackage{bm}% bold math
\begin{document} 

\title{Transition from local to global phase synchrony in small world neural network and its possible implications for epilepsy}

\author{Bethany Percha, Rhonda Dzakpasu and Micha{\l}\ \.Zochowski} 
\affiliation{Department of Physics and Biophysics Research Division\\
University of Michigan\\
Ann Arbor, MI 48109}

\begin{abstract}
Temporal correlations in the brain are thought to have very dichotomic roles. On one hand they are ubiquitously present in the healthy brain and are thought to underlie feature binding during information processing. On the other hand large scale synchronization is an underlying mechanism of epileptic seizures. In this paper we show a possible mechanism of transition to pathological coherence underlying seizure generation. We show that properties of phase synchronization in the 2-D lattice of non-identical coupled Hindmarsh-Rose neurons change radically depending on the connectivity structure. We modify the connectivity using the small world network paradigm and measure properties of phase synchronization using previously developed measure based on assessment of the distributions of relative interspike intervals \cite{mrzrd1-03}. We show that the phase synchronization undergoes a dramatic change as a function of locality of network connections from local coherence strongly dependent on the distance between two neurons to global coherence exhibiting stronger phase locking and spanning the whole network. 
\end{abstract} 
\date{\today}
\maketitle 
Epilepsy is one of the most common neurological disorders, with underlying seizures generated by indiscriminate, synchronized bursting of multiple cells in the brain \cite{deransart03}, leading to the increased level of coherence in the recorded signal between individual neurons as well as whole networks \cite{prida99,ferri04}. There is a wide range of molecular and cellular mechanisms underlying seizure generation; however, they are often linked to increased excitatory transmission mediated by NMDA, AMPA or metabotropic glutamate receptors, and a decrease in inhibitory (GABAergic) transmission, causing an imbalance between excitation and inhibition in the system \cite{dudek99}. One of the mechanisms generating the changes of the excitatory transmission under pathological conditions is axonal sprouting \cite{sutula88,cavazos91}. This mechanism involves excessive growth of excitatory processes within an area that was exposed to ischemia or physical trauma, causing (in time) generation of seizures. We hypothesize that hyperexcitability induced by sprouting could be only one of the causes of seizures and show that alteration of network structure through introduction of random long-range connectivity in the network produces relatively abrupt transition in phase coherence in the 2-D small world network (SWN) lattice of non-identical Hindmarsh-Rose models of thalamocortical neurons \cite{hindmarsh}.    

Emergence of the concept of small-world networks \cite{strogatz} has allowed for rigorous study of the properties of intermediate structured network where the connectivities are neither entirely regular not entirely random. Networks exhibiting such structure have been identified in social as well as biological systems \cite{strogatz,newman02}. Most studies have concentrated on their static properties \cite{newman99,newman00,almass02}. However, recent work has also focused on the dynamic properties of SWN, including synchronization. It has been shown that the linear stability of the synchronous state is linked to the algebraic condition of the Laplacian matrix defining network topology \cite{barahona02,hong02}. It has been reported that this synchronized state is achieved in SWN more efficiently (in terms of required network connectivity) than standard deterministic graphs, purely random graphs and ideal constructive schemes \cite{nishikawa03}. It has been also shown that the small-world networks of interconnected Hodgkin-Huxley neurons combine two features: rapid and large oscillatory response to the stimulus \cite{lago00}. Properties of self-sustained activity have also been studied in SWN of excitable neurons \cite{roxin04}.  

It has been established that periodically driven non-linear oscillators or a system of coupled non-identical oscillators can achieve phase synchronization \cite{rosenblum,rosenblum2,pikovsky,parlitz2,zhou02,pazo03}- the state in which phases are locked but the amplitudes of the signals are uncorrelated. Since in neural systems, spike generation on the level of individual cells is usually driven by the same underlying processes \cite{hh52}, although the cells' specific properties vary widely it is likely that this type of synchronization plays the most prominent role in the brain \cite{varela01}.  

Here we use a measure that was previously developed by us (\cite{mrzrd1-03}) to monitor properties of phase synchronization as a function of connectivity structure in a 2-dim lattice of non-identical, diffusively coupled Hindmarsh-Rose neurons, where the lattice has SWN structure. The measure monitors the correlations in inter-spike intervals (ISI) between the neuronal pairs (Fig 1A). This allows us to interpret the phase interdependencies of the coupled units in terms of relative inter-spike timings. The ISIs are calculated for every neuron pair in the network separately as new spikes are generated. The distributions are updated dynamically throughout the simulation. After every update, the ISI distributions are renormalized and the Shannon entropy  of the distributions is calculated. Since the distributions depend on the relative timings of spikes of both neurons in the pair, we refer to them as conditional entropies (CEs). Thus the measure provides an assessment of the instantaneous phase interdependencies between the neurons in the network without measurement of the phase itself, while the pair wise comparison of CEs allows for asymmetric measurement of phase interdependencies between any two neurons in the network. Those two characteristics of the measure make it directly applicable to experimental data.

The equations of the studied neurons are:
\begin{equation}
\begin{array}{l}
\dot{x}_{i}= y_{i}-ax_{i}^3+bx_{i}^2-z_{i}+I_{0_{i}}+\frac{\alpha}{K-1} \sum_{j, \| i,j\| \leq R}(x_{j}-x_{i})\\
\dot{y}_{i}= c-dx_{i}^2-y_{i}\\
\dot{z}_{i}= r\left[s(x_{i}-x_0)-z_{i}\right] 
\end{array}
\label{hr}
\end{equation}
Initially all neurons within radius $R$ are connected via unidirectional coupling  having strength $\alpha$. Those connections are than randomly modified with probability $P$. The $12x12$ lattice has periodic boundary conditions (i.e., torus topology); the lattice constant is set to unity. The neural parameters in above equations are: $a=1.0$, $b=3.0$, $c=1.0$, $d=5.0$, $r=0.006$, $s=4.0$, and $x_0=-1.6$; ;$K$ is the number of actual connections per neuron. The parameter $I_{0_i}$ represents the amplitude of external current applied to the $i$-th neuron and determines the frequency as well as type as the dynamical regime of the neuron (periodic, bursting and/or chaotic). The $I_{0_i}\in [2.0,3.4]$, and were generated at random, ensuring that they had non-identical properties.

The phase lag between two non-identical neurons established during phase synchronization depends on their relative properties (i.e. intrinsic frequencies). The phase of the neuron having higher frequency (higher $I_{0_i}$) will lead that of the neuron having lower frequency (Fig 1B).  The neurons, depending on the relative values of their control parameters achieve complete or phase synchronization with varying phase lag. 
\begin{figure}
\includegraphics[scale=0.7]{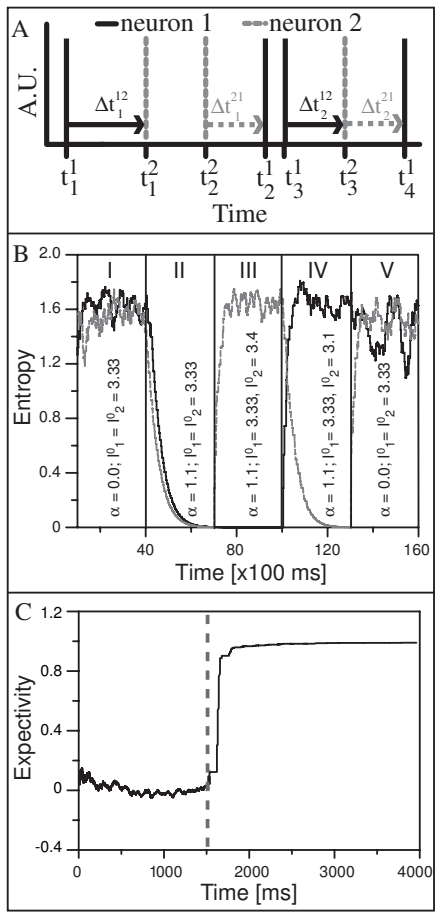}
\caption{{\bf A:} Individual distributions are updated for every neuron using the relative durations of the ISI of every neuron with respect to the other in the pair (for all the possible pairs in the network). Specifically, the ISI, $\Delta t^{ij}_m$, of the $j$-th neuron with respect to the $i$-th neuron is calculated as a time difference between the ISI timing of the $j$-th neuron with respect to the last spike of the $i$-th neuron, and conversely the ISI $\Delta t^{ji}_m$ of the $i$-th neuron is calculated as a time difference between the ISI of neuron $j$ with respect to the timing of the last spike taking place on the $j$-th neuron. The distributions are updated every time new spike is generated.{\bf B:} changes in the phase lag (as measured by CEs)in response to changes in the relative values of $I_{0_i}$ for a system two H-R neurons. Initially the  neurons are uncoupled ($\alpha = 0$) with the same control parameters ($I_{0_i}=3.3$). Both of the CEs are high (no phase synchronization is present). After $40$s the coupling is introduced ($\alpha = 1.2$). The CE of both neurons converges to zero indicating complete synchronization. After another $40$ms the value of $I_{0_2}=3.4$. The CE with respect to neuron 1 is high, whereas the other one is zero indicating phase synchronization with neuron 1 lagging behind neuron 2. After another $40$s the control parameters are modified so that $I_{0_1}=3.33$ and $I_{0_i}=3.1$ reversing that of the previous case. The CE calculated with respect to neuron 2 is high, whereas the other one is zero indicating that neuron 2 is now lagging behind neuron 1. Finally, again $\alpha=0$ and any phase relations are abolished. {\bf C:} Changes in the expectivity function for fully connected network of H-R neurons. The dashed line denotes time at which coupling was turned on ($\alpha = 2$).}
\end{figure}

Based on those results we define an expectivity function which compares the phase relations in the network to the relative properties of the neurons (the value of $I_{0_i}$):
\begin{equation}
E=\frac{1}{N(N-1)}\sum_{i,j\\,i\neq j}^N w_{ij},
\end{equation}   
where
\begin{equation}
w_{ij}=\left\{
\begin{array}{rl}
1 & \mbox{if $(S_{ij}-S_{ji})(I_{0_j}-I_{0_i})>0$}\\
-1 & \mbox{if $(S_{ij}-S_{ji})(I_{0_j}-I_{0_i})\leq 0$}
\end{array}
\right.
\end{equation} 
The expectivity function measures whether the predictions of directionality of phase lag  based on the relative values of the control parameter are in agreement with those established from assessment of pairwise differences of CEs. If the value from a given pair is predicted correctly the function is assigned the value $w_{ij}=1$, and conversely if the prediction fails $w_{ij}=-1$. Thus if there is a significant phase synchronization in the network $E\rightarrow 1$, whereas if no phase synchrony is established $E\simeq 0$ (Fig 3C).

We have used the expectivity function to measure properties of phase synchronization in a sparsely coupled 2D lattice of networked H-R neurons. The rewiring probability $P$ was varied from $0$ (full local connectivity within the radius $R$) to $1$ (random graph). The radius $R=1,2,3$ determined the connectivity fraction in the network ($0.028$, $0.083$, $0.194$ respectively). 

We have created histograms of the expectivity of all neuron pairs in the network as a function of their relative Euclidean distance on the lattice. This allowed us to infer the local as well as global properties of phase synchronization in the network. We have observed that for low values of $P$ the phase relations are preserved over short distances and the expectivities over longer distances quickly converge to zero. However, as $P$ increases global phase synchrony is achieved in the network (Fig 2). Moreover, neurons in the networks exhibiting global phase synchrony regime may achieve significantly higher degree of phase locking than that achieved even for short spatial distances in the networks with low $P$ (Fig 2). 
\begin{figure}
\includegraphics[scale=0.65]{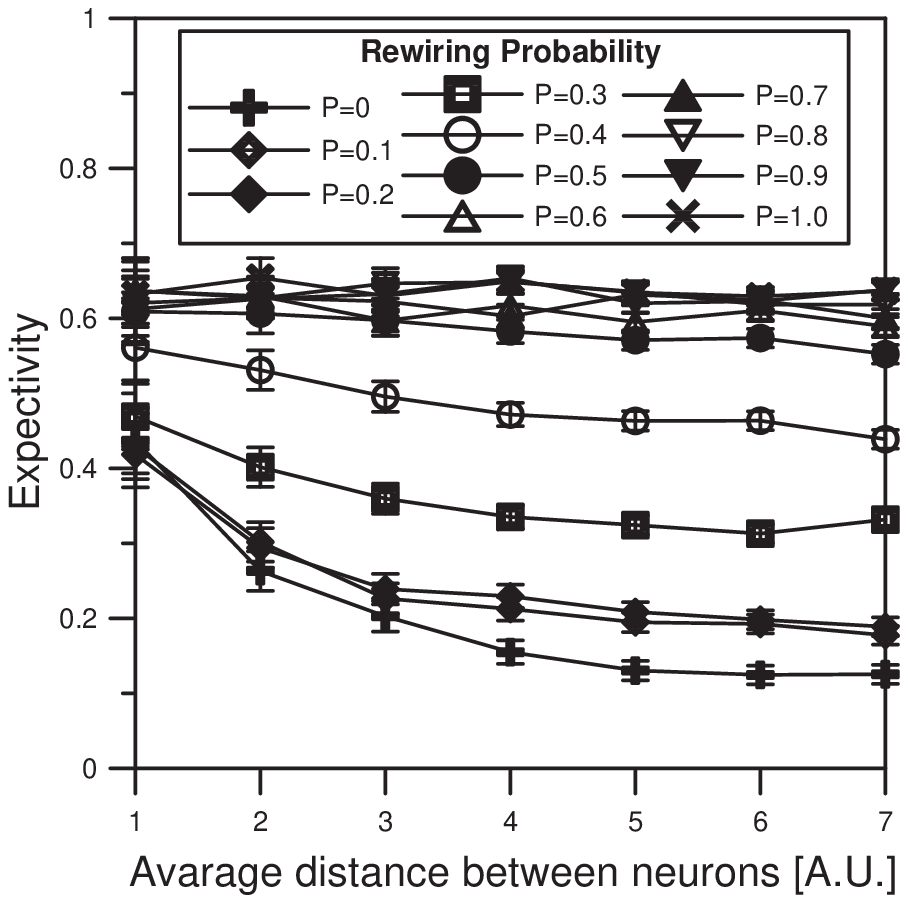}
\caption{Changes in the Expectivity as a function of neuronal distance for different rewiring probabilities. The phase synchronization remains local for low values of P (expectivity declines as a function of neuronal distance), whereas global phase synchronization is obtained for $P>0.3$. Additionally, for high values of $P$ the degree of phase locking is greater overall than that observed for local phase synchrony. The graph is formed by binning the values of expectivity for all neural pairs that have Euclidean distance within the noted distance range; $\alpha=2.0$, $R=2$. Every point on the graph is an average over 4 trials.}
\end{figure}

The abruptness of the transition from local phase synchrony to global phase synchrony depends on the connectivity as well as the coupling constant. To depict those changes, we have plotted the average decay ratio of the expectivity as a function of rewiring probability
\begin{equation}
D_{R,\alpha}(P)=\frac{E_{R,\alpha}(L=1,P=0)-E_{R,\alpha}(L=max,P)}{E_{R,\alpha}(L=1,P=0)}
\end{equation}
where $E_{R,\alpha}(L,P)$ is the expectivity averaged over all neuronal pairs having average distance $L$, computed for the network with rewiring probability $P$, radius $R$, and coupling strength $\alpha$. Positive values of the decay ($D$) indicate local phase synchrony, whereas $D\simeq 0$ indicates global phase synchrony in the network. Negative values of $D(P)$ indicate an increased level of phase locking in globally synchronous case in comparison with that of the locally synchronous case (see Fig 2).
\begin{figure}
\includegraphics[scale=0.60]{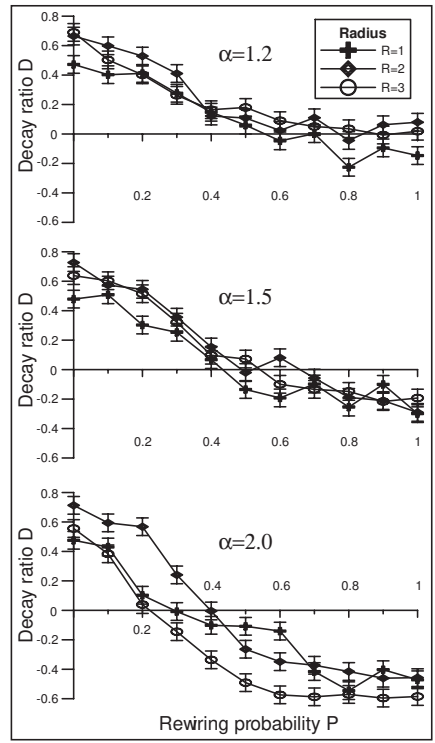}
\caption{Average synchrony decay ratio, $D$, as a function of rewiring probability P (see definition in the text). Positive values of D indicate large decay and thus local phase synchrony; for $d\simeq 0$ there is no attenuation of synchrony over distance (global synchrony state is achieved); $D>0$ indicates increased degree of phase locking within the global synchrony state.}
\end{figure}
The critical value of rewiring probability at which the transition takes place fluctuates around $P\simeq 0.3-0.4$. This coincides with the values of $P$ at which the structural clustering coefficient rapidly decays \cite{strogatz}. Moreover, it has been found that the clustering coefficient for {\it C. elegans}, an example of a completely mapped neural network is $0.28$, which corresponds to a rewiring probability of $P\simeq 0.3$ (assuming perfect SWN structure), indicating that the neural systems may form networks, where network structure lies in this critical regime between local and global synchrony. Creation of spurious glutamatergic connections in an injured region (sprouting) may cause the balance to be shifted toward global phase synchrony and thus creation of epileptic seizures. 

The increase in quality of global phase synchrony over local phase synchrony also depends on the coupling constant $alpha$. For lower values of alpha there is no significant increase in the degree of global phase synchrony over the degree of local phase synchrony observed on short distances. When the coupling is increased there is a significant enhancement in the phase synchrony in the network (up to 50\%).

In conclusion, we have applied the measure our to monitor the properties of phase synchronization in the two dimensional lattice of coupled H-R neurons having SWN connectivity. Using this measure, we have observed a transition from local phase synchrony which falls off as a function of neuronal distance, to global synchrony that is independent of this distance. We have also observed that the degree of the phase locking increases in the case of global synchrony when the coupling is strong. This effect could possibly play an important role in the emergence of epilepsy as it is known that one of the mechanisms of epileptic seizure generation is based on the sprouting of glutamatergic processes within the injured brain region.  Incidentally an additional advantage of the devised measure is the fact that it can be applied directly to experimental data. The expectivity function, which can not be assessed in the case of real data because the internal parameters of individual neurons are not known, can be substituted by pairwise calculation of the average absolute value of entropic differences between individual neurons $|\Delta S_{ij}|$. The behavior of both measures is virtually the same (Fig. 4). 
\begin{figure}
\includegraphics[scale=0.65]{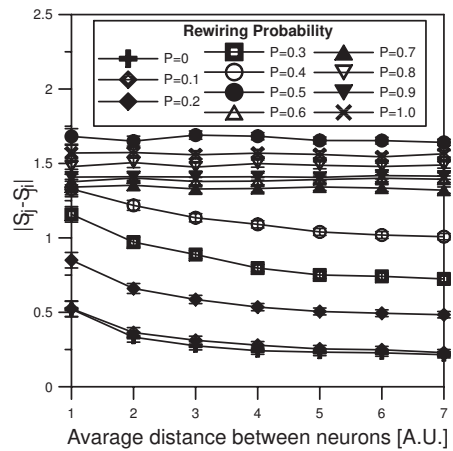}
\caption{Changes in $|\Delta S_{ij}|$ as a function of Euclidean distance between the neurons. The entropic differences exhibit the same behavior as the one observed for the expectivity function. This allows the direct application of the measure to the experimental data, where the parameters of the individual neurons can not be determined. The  $|\Delta S_{ij}|$ were calculated for the same parameters as those listed on Fig 2.}
\end{figure}

Moreover,  although it requires further investigation, it may be significant that the clustering coefficient for {\it C. elegans} implies the value of the rewiring probability that is relatively close to the observed transition point between local and global phase synchrony \cite{strogatz}, possibly indicating that the brain connectivity lies relatively close to that transition point.  

The authors would like to thank M. Newman for his comments on the work. 
%\bibliography{database,hab,epilepsy}          

\end{document}